\documentclass{article}

\usepackage{microtype}
\usepackage{graphicx}
\usepackage{booktabs} 

\usepackage{hyperref}

\usepackage[accepted]{icml2025}

\usepackage{amsmath}
\usepackage{amssymb}
\usepackage{mathtools}
\usepackage{amsthm}

\usepackage[capitalize,noabbrev]{cleveref}

\usepackage{siunitx}
\usepackage{subcaption}
\usepackage{placeins}
\usepackage{multirow}
\usepackage{bm}

\theoremstyle{plain}

\theoremstyle{definition}

\theoremstyle{remark}

\usepackage[textsize=tiny]{todonotes}

\icmltitlerunning{Neural Surrogates for Nonlinear Elastic Plates}

\begin{document}

\twocolumn[
\icmltitle{Evaluation of Neural Surrogates for Physical Modelling Synthesis of Nonlinear Elastic Plates}



\icmlsetsymbol{equal}{*}

\begin{icmlauthorlist}
\icmlauthor{Carlos De La Vega Martin}{yyy}
\icmlauthor{Rodrigo Diaz Fernandez}{yyy}
\icmlauthor{Mark Sandler}{yyy}

\end{icmlauthorlist}

\icmlaffiliation{yyy}{Centre for Digital Music, Queen Mary, University of London, London, United Kingdom}

\icmlcorrespondingauthor{Carlos De La Vega Martin}{c.delavegamartin@qmul.ac.uk}

\icmlkeywords{physical modelling, neural operators, state-space models, koopman operator}

\vskip 0.3in
]



\printAffiliationsAndNotice{}  

\begin{abstract}
Physical modelling synthesis aims to generate audio from physical simulations of vibrating structures.
Thin elastic plates are a common model for drum membranes.
Traditional numerical methods like finite differences and finite elements offer high accuracy but are computationally demanding, limiting their use in real-time audio applications.
This paper presents a comparative analysis of neural network-based approaches for solving the vibration of nonlinear elastic plates.
We evaluate several state-of-the-art models, trained on short sequences, for prediction of long sequences in an autoregressive fashion.
We show some of the limitations of these models, and why is not enough to look at the prediction error in the time domain.
We discuss the implications for real-time audio synthesis and propose future directions for improving neural approaches to model nonlinear vibration.
\end{abstract}

\section{Introduction}
\label{sec:introduction}

Modelling the dynamics of physical systems has an important role in acoustics research.
One area where they are intimately linked is physical modeling synthesis, where physics-based mathematical models of vibrating objects are used for sound synthesis.
The partial differential equations (PDEs) that govern wave propagation can be usually solved with numerical methods such as finite differences and finite elements, but these can be computationally expensive for some of the nonlinear PDEs in this context.
While these methods can provide convergence guarantees and high accuracy, they often require a compromise between speed and precision, as well as perfect knowledge of the systems physics, which can limit their applicability in real-time audio synthesis and interactive sound design.

In recent years, neural networks have emerged as an alternative approach to solving differential equations numerically.
Although training time might be longer, neural networks can amortise the cost of solving the PDE over many evaluations, and can incorporate real world data~\cite{pathakFourCastNetGlobalDatadriven2022} in a natural way.
Different techniques have been proposed to model PDE dynamics using neural networks, including neural operator methods like the Fourier Neural Operator~\cite{kovachkiNeuralOperatorLearning2023, luLearningNonlinearOperators2021}, Physics-Informed Neural Networks (PINNs)\cite{cuomoScientificMachineLearning2022}, neural ODE-based methods\cite{yinContinuousPDEDynamics2023}, and graph-based methods~\cite{brandstetterMessagePassingNeural2022}.
In parallel, advancements in State Space Models (SSMs) \cite{guModelingSequencesStructured2023} and Koopman-based architectures~\cite{mezicKoopmanOperatorGeometry2021, luschDeepLearningUniversal2018, naimanGenerativeModelingRegular2023} for long-sequence modelling have opened up new possibilities for PDE modelling in the context of music and audio.
These methods offer significant improvements over classical neural approaches, such as recurrent neural networks (RNNs), which are known to be difficult to train and prone to vanishing and exploding gradients due to backpropagation through time (BPTT)~\cite{goodfellowDeepLearning2016}.
But they still accumulate significant error~\cite{michalowskaNeuralOperatorLearning2024,diazEfficientModellingString2024,delavegamartinPhysicalModellingStiff2023} and suffer from distribution shift~\cite{,brandstetterMessagePassingNeural2022} during long rollouts.
This is is particularly relevant in the context of audio synthesis, where long sequences, on the order of hundreds of thousands of timesteps, are common.
Current models are not able to predict correctly sequences longer than a few hundred steps, usually corresponding to rollouts of  less than 10 times the training length~\cite{michalowskaNeuralOperatorLearning2024,kohlBenchmarkingAutoregressiveConditional2024,huStatespaceModelsAre2024}, despite different model-agnostic techniques proposed to mitigate the distribution shift and rollout stability issues such as the \emph{pushforward trick} and \emph{temporal bundling trick}~\cite{brandstetterMessagePassingNeural2022,brandstetterLiePointSymmetry2022}.
To our knowledge, there have been some applications of these methods in physical modelling~\cite{parkerPhysicalModelingUsing2022,delavegamartinPhysicalModellingStiff2023,diazEfficientModellingString2024}, but they don't overcome this limitations.
Memory and computational constraints also limit the length of sequences that can be processed in a single pass, specially for high-dimensional or finely discretised spatiotemporal data.
Therefore, we are interested in evaluating the performance of models trained on short blocks of the sequence, and then apply the model recurrently to generate long sequences.
As a suitable test for these methods, we consider the dynamics of a thin elastic plate described by the Berger plate model~\cite{bergerNewApproachAnalysis1954}, which is commonly used model for emulating the physics of drum membranes~\cite{fletcherPhysicsMusicalInstruments1991, avanziniEfficientSynthesisTension2012}.

The contribution of this paper is a comparison of different neural network-based methods for modelling linear and nonlinear plate vibration, with a focus on long sequences for audio synthesis.
We discuss the limitations of these methods and how some of the usual ways of evaluating them might not be enough to assess their performance in the context of audio synthesis.
The paper is structured as follows: Nonlinear plate vibration section covering the description of the PDE system and the specifics of the dataset; Methods section covering the models and training details; Results section discussing the experiments, and Conclusions.
The neural network architectures used can be found at \url{https://github.com/rodrigodzf/physmodjax}.


\section{Nonlinear plate vibration}
\label{sec:plate-vibration}

The Berger plate model~\cite{bergerNewApproachAnalysis1954} is a nonlinear PDE that describes the dynamics of a thin elastic plate undergoing moderate deflections.
The PDE can be transformed into a system of coupled ODEs by projecting over the linear eigenfunctions of the plate~\cite{avanziniEfficientSynthesisTension2012}.
This means that the nonlinearity is relatively weak for reasonable initial conditions, with the modulation mostly present shortly after the excitation.
See Appendix~\ref{app:plate} for a more detailed description of the PDE and its derivation.
This PDE can be solved efficiently using existing numerical methods, such as the Functional Transformation Method (FTM)~\cite{trautmannDigitalSoundSynthesis2003,avanziniEfficientSynthesisTension2012}, which allows us to test the performance of our models against a known solution in the context of physical modelling synthesis.

\subsection{Dataset}
We generated the dataset using the Functional Transformation Method (FTM)~\cite{trautmannDigitalSoundSynthesis2003,avanziniEfficientSynthesisTension2012}.
Each dataset consists of 100 trajectories of 1 second each, with a sampling rate of $16\si{\kilo\hertz}$.
Each trajectory is initialised with a randomly sampled gaussian-shaped velocity profile and zero displacement.
The physical parameters correspond to a rectangular mylar membrane~\cite{fletcherPhysicsMusicalInstruments1991}, tuned to a fundamental frequency of around $100\si{\hertz}$.
See Appendix~\ref{app:dataset} for further details.

\section{Models and training}
\label{sec:models}

In our experiments, we compare the performance of the Fourier Neural Operator (FNO)~\cite{kovachkiNeuralOperatorLearning2023}; a simple Koopman-based architecture (LTI) and two variants of it (LTI-MLP, LTI-SIREN) with a learned modulation~\cite{luschDeepLearningUniversal2018}, and two variants of state-space models~\cite{guEfficientlyModelingLong2022},  Linear Recurrent Unit (LRU)~\cite{orvietoResurrectingRecurrentNeural2023} model, and the S5 model~\cite{smithSimplifiedStateSpace2023}.
We train our models using a single step as input and three different output lengths, 49, 199 and 399 steps to observe the effect of \emph{temporal bundling}~\cite{brandstetterMessagePassingNeural2022}.
The best performing models are also trained using the \emph{pushforward trick}~\cite{brandstetterMessagePassingNeural2022} to try mitigate the distribution shift problem~\cite{brandstetterMessagePassingNeural2022,brandstetterLiePointSymmetry2022}.
The models are trained using the Mean Squared Error (MSE) loss function.
See Appendix~\ref{app:models} for further details on the models and their architectures, as well as training.

\subsubsection{Koopman-inspired neural methods}
\label{ssec:koopman}
The Koopman operator is a linear infinite-dimensional operator that acts on the space of observables of a dynamical system, and it can be used to study the dynamics of the system in a transformed space~\cite{koopmanHamiltonianSystemsTransformation1931}.
In discrete-time dynamical systems, a dynamical map $f: \mathcal{X} \rightarrow \mathcal{X}$ is a function that advances the system state by one time step: $\mathbf{x}_{t+1}=f\left(\mathbf{x}_t\right)$.
Consider a class of measurement functions $g \in \mathcal{G}: \mathcal{X} \rightarrow \mathbb{C}$, where $\mathcal{G}$ is a Hilbert space, producing complex-valued measurements of the system at a given state.
The Koopman operator $\mathcal{K}_f: \mathcal{G} \rightarrow \mathcal{G}$ for a dynamical map $f$ is a linear operator defined by $\mathcal{K}_f(g) \triangleq g \circ f$ \cite{koopmanHamiltonianSystemsTransformation1931}, satisfying:
\begin{equation}
  \mathcal{K}_f(g)\left(\mathbf{x}_k\right)=g \circ f\left(\mathbf{x}_k\right)=g\left(\mathbf{x}_{k+1}\right)
\end{equation}
Where $\circ$ denotes the composition of functions.
This means that the Koopman operator effectively advances the measurement functions $g$ along with the map $f$.
The infinite dimensionality of the Koopman operator makes it difficult to work with in practice, so finding a finite-dimensional  subspace or approximation is often desired~\cite{mezicKoopmanOperatorGeometry2021, mezicSpectrumKoopmanOperator2020}.
But for any given set of measurement functions $g$ identified, there are no guarantees that that they span an invariant subspace~\cite{bruntonKoopmanInvariantSubspaces2016}.
That's why most Koopman-inspired architectures~\cite{luschDeepLearningUniversal2018, huhtalaKLANNLinearisingLongTerm2024,xiongKoopmanNeuralOperator2024} usually try to learn the Koopman operator's $\mathcal{K}$ eigenfunctions, $\varphi$, and eigenvalues $\lambda$ from the data instead, as they are guaranteed to span an invariant subspace~\cite{bruntonKoopmanInvariantSubspaces2016}
\begin{equation}
\label{eq:koopman-eigen}
  \varphi\left(\mathbf{x}_{k+1}\right)=\mathcal{K} \varphi\left(\mathbf{x}_k\right)=\lambda \varphi\left(\mathbf{x}_k\right)
\end{equation}

This means that autoencoder architectures~\cite{kramerNonlinearPrincipalComponent1991} are a natural fit Koopman-based approaches~\cite{luschDeepLearningUniversal2018,naimanGenerativeModelingRegular2023}, as they can be used to implicitly learn the eigenfunctions $\varphi$ and their inverse $\varPsi \approxeq \varphi^{-1}$ as the encoder and decoder respectively.
The latent representation $\bm{x}_k$ is defined over a sequence of length $L$, where the index $k$ ranges from $0$ to $L$. It evolves through a linear time-invariant (LTI) system defined by
\begin{equation}
\bm{x}_{k+1} = \bm{\Lambda} \bm{x}_k, \quad \bm{\Lambda} = \text{diag}(\lambda_1, \lambda_2, \dots, \lambda_M) \in \mathbb{C}^{M \times M},
\end{equation}
\noindent
where $\bm{\Lambda}$ is a diagonal matrix containing complex eigenvalues $\lambda_i$.
This is similar to the approach used in the Diagonal State Space Models (DSSM)~\cite{guptaDiagonalStateSpaces2022}, making it very computationally efficient thanks to the use of the \emph{parallel scan} algorithm~\cite{blellochPrefixSumsTheir2004}.


\section{Results}

To evaluate the performance of the models, we use two different tasks: single block prediction and long sequence prediction.

For the single block prediction task, we evaluate the models' performance generating a single block of the sequence, which is how the models were trained.
For ease of comparison MAE and MSE are normalised with respect to the corresponding ($L_1$ or $L_2$) norm of the ground truth. The results are shown in Table~\ref{tab:mse_mae}.
The evaluation samples are drawn from the test set, 100 samples apart for all trajectories.
For most architectures, the error scales roughly linearly with the output length in the case of MSE, but MAE  .
The best performing model, LRU, achieves almost an order of magnitude lower error for the shortest output length, but the advantage is reduced for longer sequences.

\begin{table}[!hb]
    \tiny
    \centering
    \resizebox{0.40\textwidth}{!}{
        \begin{tabular}{@{}llll@{}}
            \toprule
            Model     & Steps & Relative MSE   & Relative MAE   \\ \midrule
            FNO       & 49    & 0.0030(0.0000) & 0.0423(0.0008) \\
                      & 199   & 0.0133(0.0008) & 0.0864(0.0039) \\
                      & 399   & 0.0306(0.0018) & 0.1358(0.0074) \\ \midrule
            LTI       & 49    & 0.0059(0.0016) & 0.0640(0.0098) \\
                      & 199   & 0.0230(0.0036) & 0.1066(0.0099) \\
                      & 399   & 0.0354(0.0018) & 0.1324(0.0052) \\ \midrule
            LTI-MLP   & 49    & 0.0064(0.0006) & 0.0721(0.0033) \\
                      & 199   & 0.0141(0.0024) & 0.0929(0.0064) \\
                      & 399   & 0.0188(0.0001) & 0.0933(0.0038) \\ \midrule
            LTI-SIREN & 49    & 0.0050(0.0007) & 0.0587(0.0041) \\
                      & 199   & 0.0205(0.0013) & 0.1002(0.0043) \\
                      & 399   & 0.0333(0.0031) & 0.1215(0.0043) \\ \midrule
            S5        & 49    & 0.0054(0.0014) & 0.0470(0.0095) \\
                      & 199   & 0.0228(0.0018) & 0.0817(0.0075) \\
                      & 399   & 0.0456(0.0021) & 0.1027(0.0038) \\ \midrule
            LRU       & 49    & \textbf{0.0006(0.0000)} & \textbf{0.0178(0.0006)} \\
                      & 199   & \textbf{0.0056(0.0002)} & \textbf{0.0384(0.0038)} \\
                      & 399   & \textbf{0.0144(0.0010)} & \textbf{0.0500(0.0012)} \\ \midrule
            LRU + PF  & 49    & 0.0012(0.0003) & 0.0274(0.0060) \\
                      & 399   & 0.0280(0.0016) & 0.0758(0.0019) \\ \bottomrule
            \end{tabular}
    }
    \caption{Relative MSE and MAE across different models and steps. The numbers in parentheses represent the standard deviation across three seeds. Lowest values per case are in bold.}
    \label{tab:mse_mae}
\end{table}

\begin{figure}
    \centering
    \includegraphics[width=0.35\textwidth]{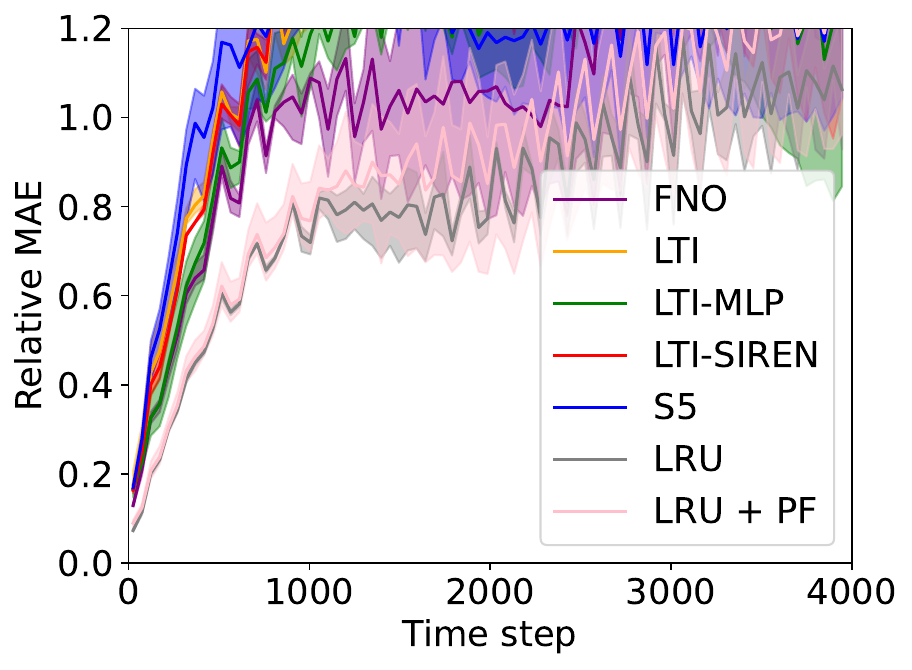}
    \caption{Relative MAE per block during the first 250 \si{\milli\second} for different models and output legth 49. The shaded area represents the standard deviation across three seeds. PF stands for pushforward trick.}
    \label{fig:mae_rel_per_slice}
\end{figure}

In the long sequence prediction task, we evaluate the models' performance generating up to 4000 steps recurrently.
The models are given a single step as input and the last step of the output is used as input for the next prediction of the next block.
The normalised MAE is calculated for each block with the results for 49 and 399 steps of output for shown in Fig.~\ref{fig:mae_rel_per_slice}.

\begin{figure}[ht]
    \centering
    \begin{subfigure}{0.4\textwidth}
        \includegraphics[width=\textwidth]{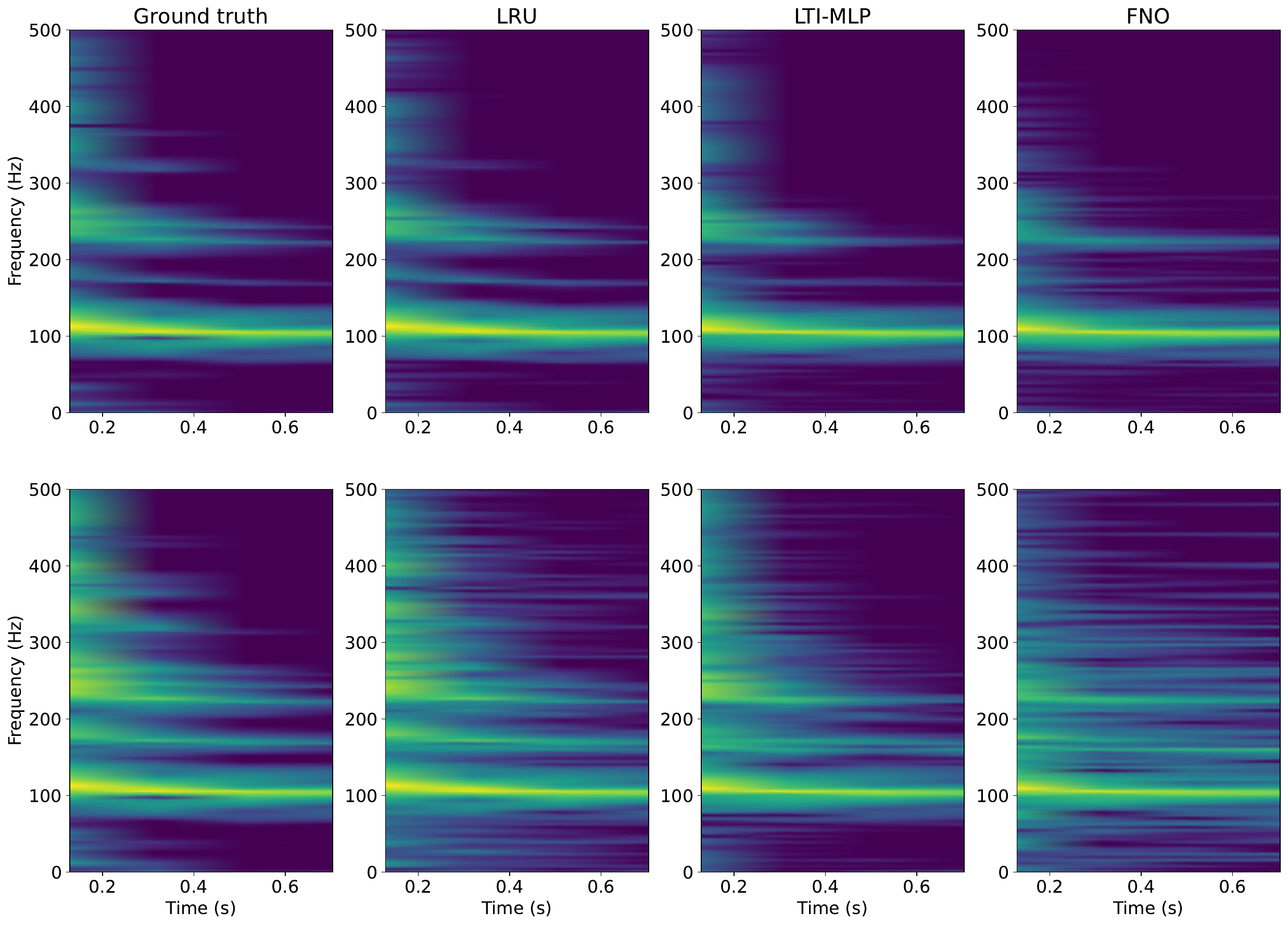}
    \caption{399 steps output.}
    \label{ffig:spectrograms-399}
    \end{subfigure}
    \begin{subfigure}{0.4\textwidth}
        \includegraphics[width=\textwidth]{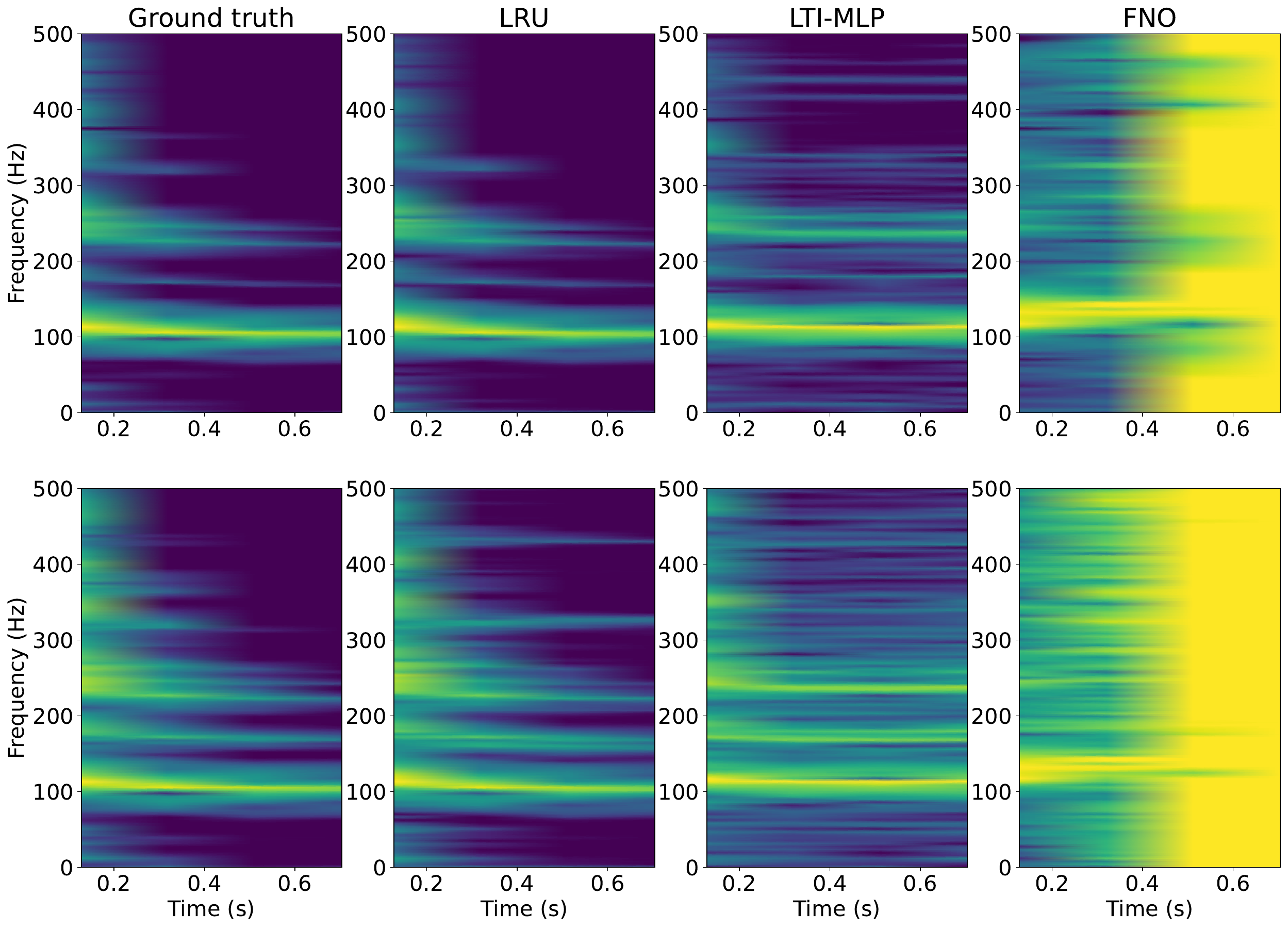}
    \caption{49 steps output.}
    \label{fig:spectrograms-49}
    \end{subfigure}
    \caption{Spectrograms for the best performing model in different classes. Top is for a point near the center of the plate, bottom is for a point near the edge.}
    \label{fig:spectrograms}
\end{figure}

\begin{figure}[ht]
    \centering
    \begin{subfigure}{0.4\textwidth}
        \includegraphics[width=\textwidth]{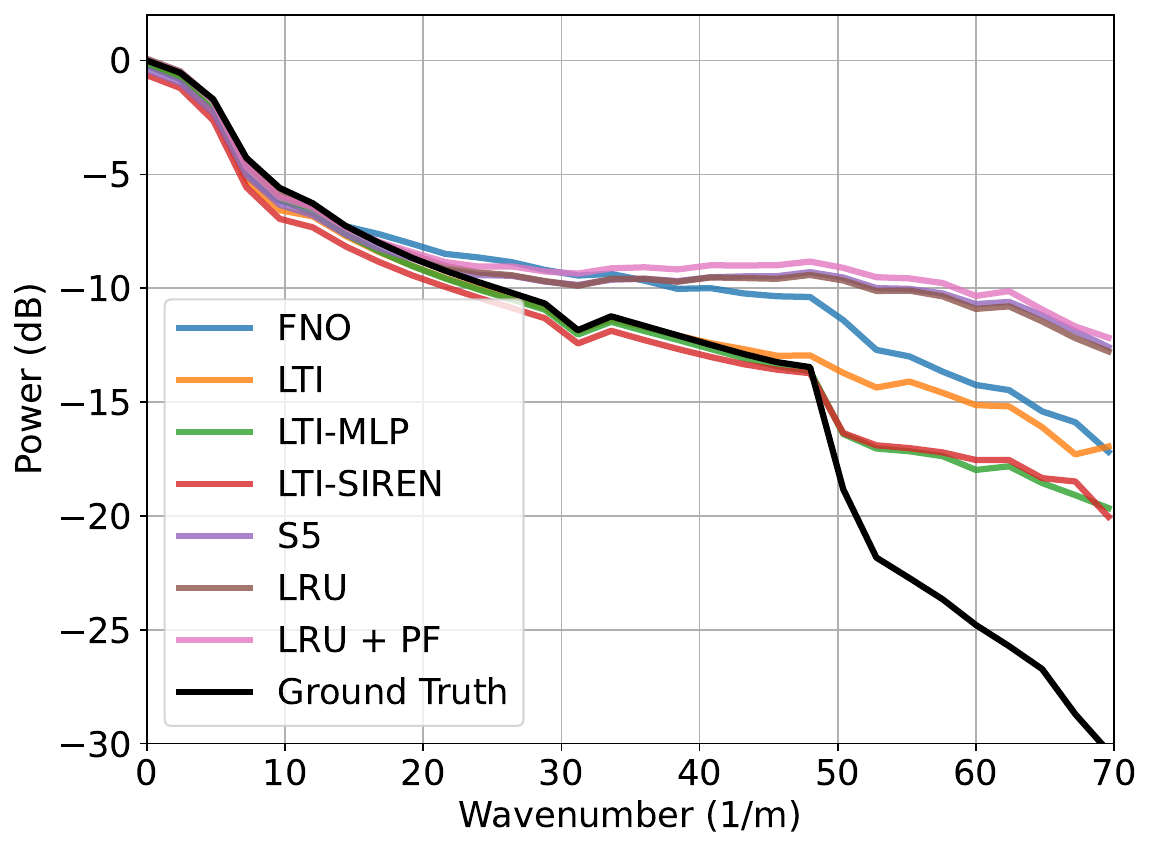}
    \caption{399 steps output.}
    \label{fig:spatial_power_spectrum_399}
    \end{subfigure}
    \begin{subfigure}{0.4\textwidth}
        \includegraphics[width=\textwidth]{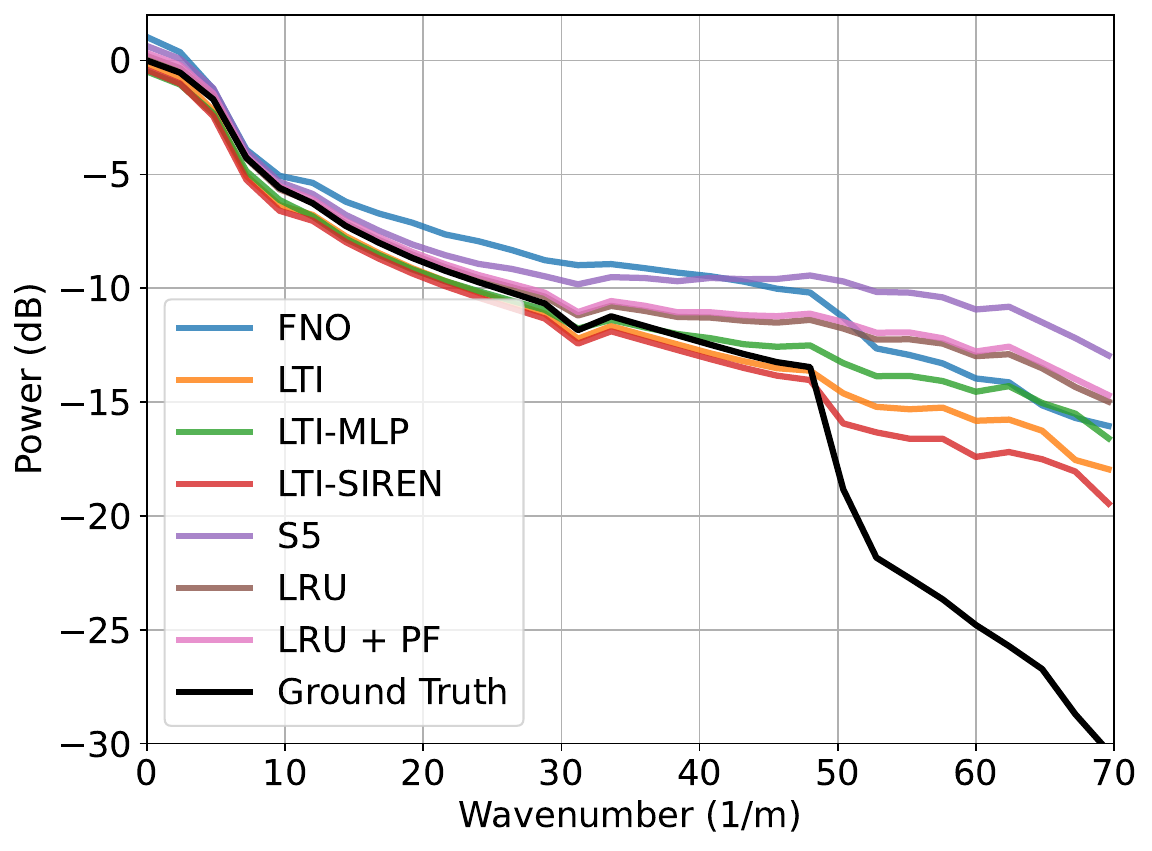}
    \caption{49 steps output.}
    \label{fig:spatial_power_spectrum_49}
    \end{subfigure}
    \caption{Radial spatial power spectrum of the predictions for the best performing model in different classes, averaged over the first 4000 samples of the trajectory.}
    \label{fig:spatial_power_spectrum}
\end{figure}

As it can be seen the results for the single block prediction task do not translate to the long sequence prediction task, with the models struggling particularly at the beginning of the sequence.
The LRU model performs the best in the long sequence prediction task, but training with the pushforward trick does not meaningfully improve the results.
When looking at the spectrograms of the predictions for the full second (16000 samples) of the two best performing models of several classes, we can see that the behaviour of the models is quite different between the center and the edges of the plate, and in general the models tend to add more energy higher in the spectrum, as shown in Fig.~\ref{fig:spectrograms}.
When looking at the time-averaged radial spatial power spectrum of the predictions, which is shown in Fig.~\ref{fig:spatial_power_spectrum} we can also see an excess of energy in the higher wavenumbers, specially beyond 50 \si{\metre^{-1}}, where due to the minimum standard deviation of the gaussian-shaped initial conditions (see Appendix~\ref{app:dataset}) we would not expect to see much energy.
The case of the FNO is particularly interesting, as despite having very similar spatial frequency content at the 2 plotted sequence lengths, the temporal frequency content is very different, even becoming unstable as see in Fig.~\ref{fig:spectrograms-49}.
These discrepancies between the spatial and temporal spectral domains behaviour suggest that the models are not learning coherent spatiotemporal modes with a global behaviour.
Despite the performance of the LTI-based models in the single block prediction task being worse, their spatial spectra can be more accurate than the LRU.
In general, LTI-based models perform better on the spatial and temporal frequency domains when using longer sequences.

\section{Conclusions}
\label{sec:conclusion}

In this paper we have presented a comparison of neural network-based methods for modelling the dynamics of nonlinear plate vibrations in the context of physical modelling synthesis.
Although some models perform well in the single block prediction task, none of the models are able to capture the dynamics of the plate in the long sequence prediction task, making them not suitable in the current state.
These results show the need for more robust evaluation metrics for these models, specially in the context of audio synthesis, where the frequency content is extremely important.
Another important aspect is to notice that despite Koopman theory~\cite{koopmanHamiltonianSystemsTransformation1931} being claimed as a theoretical basis for many  models, in an autoencoder setting we are usually compressing the data into a low-dimensional latent space, not expanding it into a high-dimensional space as Koopman theory would suggest.
This might be part of the reasons why the frequency content in the spatial and temporal frequency domains is not neccesarily correlated, which woud be expected if we were finding true eigenfunctions.
In their current state, using neural approaches for the solution of time-dependent PDEs relevant to physical modelling synthesis is not the most efficient approach, but thorough evaluation of these methods can reveal areas of possible improvement.
The ability to work from a data-driven perspective opens the doors for applications where Finite Element or Finite Difference methods are not suitable, making relevant the characterisation of these methods as a whole.

\section*{Acknowledgements}
Carlos De La Vega Martin and Rodrigo Diaz Fernandez are research students at the UKRI Centre for Doctoral Training in Artificial Intelligence and Music, supported by UK Research and Innovation [grant number EP/S022694/1].
This research utilised Queen Mary's Apocrita HPC facility, supported by QMUL Research-IT. http://doi.org/10.5281/zenodo.438045

\FloatBarrier
\newpage
\bibliography{paper-icml2025-mlaworkshop}
\bibliographystyle{icml2025}

\newpage
\appendix
\section{Berger Plate Theory}
\label{app:plate}

Considering Euler-Bernoulli's beam theory~\cite{morseTheoreticalAcoustics1968} for the stiffness, the displacement $u(\boldsymbol{x},t), \boldsymbol{x} \in \mathcal{S} \subset \mathbb{R}^2$ of a thin elastic plate at time $t$ can be expressed as,
\begin{align}
D \nabla^4 u +\rho_{2} \frac{\partial^2 u}{\partial t^2} - T(u) \nabla^2 u + d_{1} \frac{\partial u}{\partial t} + d_{3} \frac{\partial \nabla^2 u}{\partial t} = f^{(ext)},
\label{eq:plate}
\end{align}
\noindent
where $\nabla^2$ is the Laplace operator and $\nabla^4$ is the biharmonic operator.
$\rho_{2}$ stands for the material density ($\si{\kilo\gram}/\si{\meter}^2$), D is the bending stiffness ($\si{\newton} \times \si{\meter}$), $T$ is the tension ($\si{\newton}/\si{\meter}$), $d_1$ is the frequency dependent damping ($\si{\kilo\gram}/\si{\meter}^2\si{\second}$) and $d_3$ is the frequency independent damping ($\si{\kilo\gram} \times \si{\meter}^2/\si{\second}$).
$f^{(ext)}$ is the external forcing of the system ($\si{\newton}/\si{\meter}^{2}$).
Eq.~\ref{eq:plate}, together with appropriate boundary conditions over $\partial\mathcal{S}$, and initial conditions $u(\boldsymbol{x},0)$ and $\frac{\partial u}{\partial t}(\boldsymbol{x},0)$, fully describe the system.
The tension can be divided into a static and a state-dependent correction $T(u) = T_0 + T_{NL}(u)$, and the solution can be expressed in terms of the linear modal shapes, with their corresponding eigenfrequencies instantaneoulsy modulated by the tension induced by the deformation of the membrane~\cite{bergerNewApproachAnalysis1954}.
In the linear case, $T_{NL} = 0$, we can express the solution as a sum of spatiotemporal modes, where the spatial and temporal dependencies have been decoupled,
\begin{equation}    
u(\boldsymbol{x},t) = \sum_\eta \frac{\bar{u}_\eta(t) K_\eta(\boldsymbol{x})}{\lVert K_{\eta}(\boldsymbol{x}) \rVert_{2}^{2}}, \eta \in \mathbb{N}^d,
\end{equation}
\noindent 
and the modal shapes obey,
\begin{equation}    
\nabla^2 K_\eta(\boldsymbol{x}) =-\lambda_{\eta} K_\eta(\boldsymbol{x}).
\end{equation}
\noindent 
Transforming to modal coordinates $\bar{u}_{\eta}(t)$, the PDE can be expressed as a system of ODEs,
\begin{align}
\rho_{d} \ddot{\bar{u}}_{\eta} + \left(d_3 \lambda_{\eta}+d_1\right) \dot{\bar{u}}_{\eta}+ \left(\lambda_{\eta} \left(\lambda_{\eta}D +T_0\right)\right) \bar{u}_{\eta} = \bar{f}_{\eta}.
\end{align}
\noindent
In the nonlinear case, a common simplifying assumption is that the tension is homogeneous, and therefore it only depends on the global state, which is usually called the Berger plate theory~\cite{bergerNewApproachAnalysis1954}.
\begin{equation}    
T_{N L}(u)=C_{N L} \frac{S(u)-S_0}{S_0} \simeq \frac{1}{2} \frac{C_{N L}}{S_0} \int_{\mathcal{S}}\|\nabla u\|^2 d \boldsymbol{x},
\end{equation}
\noindent
where $S(u)$ is the extension (area of the membrane) and $C_{N L}$ is a coefficient that depends on the geometry of the problem \cite{avanziniEfficientSynthesisTension2012}.
Under these assumptions $T_{NL}$ can be seen a forcing term, with the spatial mode shapes remaining the same.
This forcing term can be projected into modal coordinates,
\begin{align}
\bar{f}^{(tm)}_{\eta}(u, \bar{u}) & = -\lambda_{\eta} T_{N L} (u) \bar{u}_{\eta}(t) \\
& = -\lambda_{\eta} \frac{1}{2}\frac{C_{N L}}{S_0} \left[\sum_{\tilde{\eta}} \frac{\lambda_{\tilde{\eta}}\bar{u}_{\tilde{\eta}}^2 (t)}{\lVert K_{\tilde{\eta}} \rVert_{2}^{2}} \right] \bar{u}_{\eta}(t),
\end{align}
\noindent
and given a set of transformed initial conditions $\bar{u}_0, \dot{\bar{u}}_0$ these ODEs can then be integrated numerically using various methods, such as the Runge-Kutta method (RK4), to obtain the time evolution of the amplitudes $\bar{u}_{\eta}(t)$.
The resulting solution can be used to reconstruct the displacement $u(\boldsymbol{x},t)$ and velocity $\dot{u}(\boldsymbol{x},t)$.

\section{Dataset details}
\label{app:dataset}
The maximum initial velocity is in the range of $5$ to $25 \si{\meter}/\si{\second}$.
The PDE parameters are $\rho_{2} = 0.2622\si{\kilo\gram}/\si{\meter}^2$, $D = 2.198 \times 10^{-3}\si{\newton} \times \si{\meter}$, $T_0 = 800 \si{\newton}/\si{\meter}$ , $d_1 = 0.5\si{\kilo\gram}/\si{\meter}^2\si{\second}$, $d_3 = 0.005\si{\kilo\gram} \times \si{\meter}^2/\si{\second}$.
The rectangular domain is $40 \times 36 \si{\centi\metre}$, discretised on a grid of $N_x = 41$ and $N_y = 37$ points, therefore  $\Delta x = \Delta y = 0.01 \si{\meter}$.
The dataset consists of 100 trajectories of 1 second each, with a sampling rate of $16\si{\kilo\hertz}$, with simply-supported boundary conditions.
The solver uses uses the \emph{Functional Transformation Method} (FTM)~\cite{trautmannDigitalSoundSynthesis2003,avanziniEfficientSynthesisTension2012} to solve the PDE, with the number of modes used for the projection being 15 in each dimension, (i.e. 225 modes in total).
The gaussian-shaped initial velocity profiles have an standard deviation between $0.02$ and $0.1 \si{\metre}$.

\section{Model details}
\label{app:models}

We implement all models for this work using JAX.
The input for our models consists of the position and velocity information at a regular grid of points over the at regularly sampled time intervals.

\subsection{Fourier Neural Operator}
The Fourier Neural Operator (FNO) implementation is adapted for JAX from the implementations in~\cite{kovachkiNeuralOperatorLearning2023,parkerPhysicalModelingUsing2022}.
A single \textit{fourier} layer  comprises a projection onto the Fourier series, truncation, and multiplication of the Fourier coefficients by trainable parameters.
Then the result is padded before applying the inverse Fourier transform, applying a nonlinear point-wise transformation and adding a skip connection.
Our FNO model stacks eight (8) of these layers, using 20 modes in each dimension for the truncation, and 32 hidden channels.

\subsection{State Space Models}
\label{ssec:ssm}
State Space Models (SSMs) provide a structured approach for modeling dynamic systems with control inputs, in  a sequence-to-sequence fashion.
The SSMs strategy involves layering multiple linear time-invariant systems and introducing nonlinearities only after computing linear trajectories.
Diagonalisation of these linear rollouts makes their application much more compute efficient than usual recursive architectures~\cite{guptaDiagonalStateSpaces2022,orvietoResurrectingRecurrentNeural2023}.
To adapt the models to our specific problem, we modify the original S5\footnote{\url{https://github.com/lindermanlab/S5}} and an unofficial LRU\footnote{\url{https://github.com/NicolasZucchet/minimal-LRU}} implementations by introducing an additional initial layer that generates a sequence from the initial conditions.

\subsection{LTI models (Koopman-inspired)}
\label{ssec:lti}
\begin{figure*}[ht]
    \centering
    \begin{subfigure}{\columnwidth}
    \fontsize{7.0pt}{10pt}\selectfont
    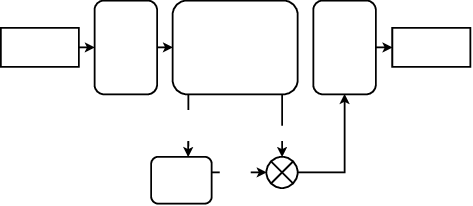
    \label{fig:1a}
    \end{subfigure}
    \begin{subfigure}{\columnwidth}
    \fontsize{7.0pt}{10pt}\selectfont
    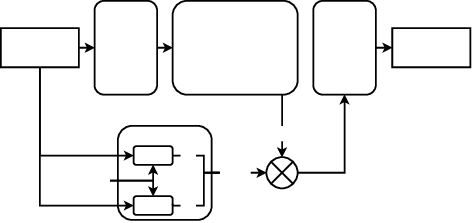
    \label{fig:1b}
    \end{subfigure}
    \caption{Architecture of LTI-MLP showing linear dynamics post-processing, involving an MLP using the square of eigenvalues radii as input and providing output for element-wise multiplication with the state sequence. On the right, we show LTI-SIREN, a similar architecture which features a modulated MLP with sinusoidal activations.}
    \label{fig:architectures}
  \end{figure*}

All of the \textit{LTI} models use a 4 layer MLP for the encoder and a single layer for the decoder, which was found to perform better than a MLP-based decoder.
The \textit{LTI-MLP} and \textit{LTI-SIREN} models use a time-varying module that modulates the eigenvalues of the linear dynamics, as shown in Fig.~\ref{fig:architectures}.
Optimisation of the diagonal entries of $\Lambda$ can lead the unbounded exponential growth if the eigenvalues take on a positive value for the real part.
To address this issue, we parametrise the eigenvalues with a clipped polar representation, with the radius in the range $[0, 1]$ to ensure that the eigenvalues remain stable during training by enforcing negativity.

\subsection{Training details}
\label{ssec:training}
The dataset is divided into 80 trajectories for training 10 for validation and 10 reserved for test.
Prior to training, we normalise the dataset to to have a standard deviation of 1.
The normalisation is calculated over the training set only, and the same normalisation is applied to the validation and test sets.
For each epoch a batch of 25 blocks with the appropriate number of steps is sampled from each trajectory in the training set.
We optimise our model using the AdamW optimiser with a fixed learning rate of 0.0001.
For the LRU and S5 models, the SSM block uses a smaller learning rate of 0.000025.
Gradient clipping is used to stabilse training, with a maximum norm of 1.0.
The training process is carried out for 5000 epochs with early stopping, with the criterion being the normalised MAE on the validation set, which is evaluated every 50 epochs.

\end{document}